\magnification=1200
\baselineskip 20pt
\def\prref{\par\noindent\hangindent=0.3cm\hangafter=1}

\def\pmbNo1{\setbox0=\hbox{$No1$}%
  \kern-0.25em\copy0\kern-\wd0
  \kern.05em\copy0\kern-\wd0
  \kern-0.025em\raise.0433em\box0}

\def\prref{\par\noindent\hangindent=0.3cm\hangafter=1}

\def \cntl{\centerline}
\def \etal {{\it et al.}}
\def \cf {{\it cf.}}
\def \eg {{\it eg.}}

\def \ie {{\it i.e.}}

\def \br {{ \bf r}}

\def \bk {{ \bf k}}

\def\iras{IRAS}
\def\pscz{PSCZ}
\def \bs {{ \bf s}}
\def \bd {{ \bf d}}
\def \bx {{ \bf x}}
\def \vd {{ \bf d}}
\def \bR {{ \bf R}}

\def \bepsilon {{\bf \epsilon}}
\def \WF {^{\rm WF}}
\def \CR {^{\rm CR}}

\def\rd{{\rm d}}

\def \apj {{\it Ap. J.}}
\def \ApJS {{\it Ap. J. Suppl.}}
\def \ApJLet {{\it Ap. J. Lett.}}
\def \MNRAS  {{\it M.N.R.A.S.}}
\def \mnras  {{\it M.N.R.A.S.}}

\def \vjec {\vfill\eject}

\def \r0p{ r{_0^\prime}}
\def\prior{{\it prior}}

\def\myfig#1#2#3#4#5{
  \midinsert
     \vskip #2 true cm \vskip 1.3 true cm
     #3
     \vskip #4 true cm
     {\baselineskip 11pt
       \rightskip=0.8 true cm \leftskip=0.8 true cm
       \par\noindent{\bf Figure #1:}  #5 \par
       \rightskip=0 true cm \leftskip=0 true cm
     }
     \vskip -0.3 true cm
  \endinsert
}

\nopagenumbers
\cntl{\bf Non-linear Constrained Realizations of the Large Scale Structure}
\vskip 8pt
\cntl{\bf  V. Bistolas\footnote{$^1$}{E-mail: bisto@vms.huji.ac.il} and
Y. Hoffman\footnote{$^2$}{E-mail: hoffman@vms.huji.ac.il}
}
\vskip 8pt
\cntl{ Racah Institute of Physics, The Hebrew University }
\cntl{Jerusalem, Israel}
\vskip 8pt
\cntl{\bf ABSTRACT}
\bigskip

The linear algorithm of the Wiener filter and constrained realizations (CRs) of Gaussian
random fields is extended here to perform non-linear CRs. The
procedure consists of: (1) Using   low resolution data to constrain  a  high
resolution realization of the  underlying field, as if the linear theory is valid; (2)
Taking the linear CR backwards in time, by the linear theory, to set initial conditions
for    N-body simulations; (3) Forwarding the field  in time by an N-body code. An
intermediate step is introduced to `linearize' the low resolution data.

The non-linear CR can be applied to any observational data set that is
quasi-linearly related to the underlying field.  Here it is applied to the \iras\ 1.2Jy
catalog using 846 data  points within a sphere of $6000 \, km/s$, to reconstruct the full
non-linear large scale structure of our `local' universe. The method is tested
against mock IRAS surveys, taken from random non-linear realizations. A detailed analysis
of the reconstructed non-linear structure is presented.

{\it Subject headings:} cosmology: large-scale structure of universe, methods: statistical

\vjec
\cntl{\bf I. Introduction}
\bigskip

\footline= {\hss\tenrm\folio\hss}
\pageno=2
In the standard model of cosmology galaxies and the large scale structure of
the universe form out of a  random perturbation field {\it via}
gravitational instability. It is assumed that  the primordial
perturbation field constitutes a random homogeneous and isotropic Gaussian field and that
on relevant scales its amplitude  is small, hence its dynamics is described by
the linear theory of gravitational instability  (\cf\ Peebles 1980). The theoretical
study of structure formation has been a major effort of modern cosmology
(\cf\ Padmanabhan 1993). On the observational side, the large scale structure has been
studied mostly by means of red-shift surveys (\cf\  Strauss and Willick 1995) and
peculiar velocities (\cf\  Dekel 1994). A method
for the reconstruction of the underlying dynamical (density and velocity) fields from a
given  observational data base is presented here.

The problem of recovering the underlying field from    given observations,
which by their nature are incomplete and have a finite accuracy and resolution, is one
often encountered in many branches of physics and astronomy. It has been shown that
for a random Gaussian field  an optimal estimator of the underlying field is given by
a minimal variance solution (Zaroubi, Hoffman, Fisher and Lahav 1995; ZHFL), known also
as the Wiener filter (hereafter WF, Wiener 1949, Press \etal\ 1992). This approach is
based on the assumed  knowledge of the second moments of the random field. These
moments, also known as co-variance matrices, are to be deduced directly from the data,
or to be calculated from an assumed model,  the so-called \prior. Within the
framework of Gaussian fields the WF coincides with the Bayesian {\it  posterior}  and
the maximum entropy estimations (ZHFL). Indeed, in the cosmological case where on large
enough scales the linear theory applies and the (over)density and velocity fields are
Gaussian the WF is the optimal tool for the reconstruction of the large scale structure.
This is further complemented by the algorithm of constrained realizations  (CRs) of
Gaussian fields (Hoffman and Ribak 1991) to create Monte Carlo simulations of the
residual from this optimal estimation. This combined WF/CR approach has been applied
recently to a variety of cosmological data bases in an effort to recover the large
scale structure. This includes the analysis of the COBE/DMR data (Bunn \etal\ 1994,
Bunn, Hoffman and Silk 1996), the analysis of the velocity potential (Ganon and Hoffman
1993), the reconstruction of the density field  (Hoffman 1993, 1994, Lahav 1993, 1994,
Lahav \etal\ 1994, Webster, Lahav and Fisher 1996) and the peculiar velocity field (Fisher
\etal\ 1995a) from the \iras\ redshift survey (Fisher \etal\ 1993). It has also
been recently  applied to the MARKIII peculiar velocities (Willick \etal\
1995) to  reconstruct the underlying dynamical fields (Zaroubi, Hoffman and Dekel 1996).

A major limitation of the WF/CR approach is that it applies only in the linear regime.
Yet, on
small scales the perturbations are not small and the full non-linear gravitational
instability theory has to be used. Here  the WF/CR method  is extended to
the quasi-linear regime, and a new algorithm of non-linear constrained realizations
(NLCRs) is presented. The general method of WF and CRs  and its modification to the
case of quasi-linear data is presented in \S II. The method is tested against N-body
simulations  and its application to the \iras\ 1.2Jy catalog (Fisher \etal\
1995b) is given in \S III. The
NLCRs of our 'local' universe  are presented in \S IV and  a short discussion
(\S V) concludes the paper.

\bigskip
\cntl{\bf II. Non Linear Constrained Realizations}
\bigskip

{\bf a. Linear Theory}

\medskip

The general WF/CR method has been fully described in ZHFL and only a very short
outline of it is presented here.  Consider the case of a set of observations performed
on an underlying random field (with $N$ degrees of freedom) $\bs=\{s_1,...,s_N\}$
yielding $M$ observables, $\bd=\{d_1,...,d_M\}$. Here, only measurements that can be
modeled as a linear convolution or mapping on the field are considered. The act of
observation is represented by
$$
\bd = \bR \bs + \bepsilon,  \eqno(1)
$$
where $\bR$ is a linear operator which represent a point spread function
and $ \bepsilon = \{\epsilon_i,...,\epsilon_M\}$ are the statistical errors. Here the
notion of a point spread function is extended to include any linear operation that
relates the measurements to the underlying field. The WF  estimator is (ZHFL):
$$
\bs{\WF} = \Bigl<\bs \,\vd^\dagger\Bigr>\Bigl<\vd\,\vd^\dagger\Bigr>^{-1}
\vd \quad .\eqno(2)
$$
Here, $\Bigl<  ... \Bigr>$ represents an ensemble  average and
$\Bigl<\bs \, \vd^\dagger\Bigr>$ is the cross-correlation matrix of the data and the
underlying field. The data auto correlation matrix is
$$
\Bigl<\bd \, \vd^\dagger\Bigr> =\bR \Bigl<\bs \, \bs^\dagger\Bigr> \bR^\dagger +
\Bigl<\bepsilon \, \bepsilon^\dagger\Bigr>,  \eqno(3)
$$
where the second term represents the statistical errors, \ie\    shot noise.

In the case of a random Gaussian field, the WF estimator coincides with the conditional mean field given the data. A  CR of the random residual from this mean field is obtained by
creating an unconstrained realization of the underlying field ($ \tilde
\bs$) and the errors ($ \tilde \bepsilon$), and `observing' it the same way the actual
universe is observed. Namely, a mock data base is created by:
$$
\tilde \bd = \bR \tilde \bs + \tilde  \bepsilon,  \eqno(4)
$$
A CR is then simply obtained by (Hoffman and Ribak 1991):
$$
\bs\CR = \tilde \bs + \Bigl<\bs \,\vd^\dagger\Bigr>\Bigl<\vd\,\vd^\dagger\Bigr>^{-1}
\bigl( \bd   - \tilde \bd \bigr). \eqno(5)
$$

{\bf b. Covariance Matrices and Shot Noise}

\medskip

The WF/CR and the NLCR presented here can be used with any data base whose relation to
the underlying field can be modeled by Eq. 1. Thus for example, observations of the velocity field
can be used to reconstruct the density field and {\it vice versa}. The concrete case
studied here is the reconstruction of the continuous density field from a
discrete galaxy catalog within a given volume subject to certain
selection criteria. Here we assume the galaxy distribution in configuration space is
known and red-shift distortions are ignored. The formalism of WF/CR can be expressed
in any functional representation, such as Fourier or spherical harmonics/Bessel
functions. The choice of the particular representation is usually dictated by the
geometry of the observations. In the case of a full sky coverage and radial selection
function the obvious choice is the spherical harmonics/Bessel basis (Fisher \etal\
1995a, Webster, Lahav and Fisher 1996). However, in the case of an incomplete sky survey
the so-called zone of avoidance (ZOA) couples the spherical harmonics and this results in a
complicated covariance matrix. For the general case we choose here to use the
configuration space representation and the field is evaluated on a Cartesian grid. The
estimation of the continuous field is done by smoothing, \ie\ the convolution of the
discrete galaxy distribution with a certain kernel. The dicreteness of
the galaxies introduces shot-noise errors and the smoothing procedure
cause these errors to be correlated to each other.
The smoothing kernel depends on the nature of the data and
is determined by a compromise between  two conflicting considerations, namely high
resolution and low noise level. A high resolution is achieved by using a narrow
smoothing kernel, and the noise level is reduced by widening the kernel. Here a
Gaussian filter is used with two smoothing length radii, $R_L$ and $R_S$. The data is
smoothed on a large scale $R_L$ and high resolution CRs are created on the scale
$R_S$, thus $R_L > R_S$.

An estimator of the fractional over-density at the point $\br_\alpha$ is given by
$$
\Delta_\alpha = \Delta(\br_\alpha) = {1 \over \bar n (2 \pi R{^2_L})^{3/2}}
  \sum_{\rm gal}
 {1\over \phi(r_{\rm gal})   }
 \exp\bigl( - { (\br_\alpha - \br_{\rm gal})^2 \over {2 R{^2_L}}  } \bigr)
\  - \ 1,     \eqno(6)
$$
where $ \bar n$ is the mean number density of the galaxies and $\phi(r)$ is the
data selection function. The data autocorrelation function is written as $ \Bigl<
\Delta_\alpha \Delta_\beta\Bigr> = \xi_{\alpha \beta} + \sigma_{\alpha \beta} $. The
first term is just the autocorrelation function of the smoothed field ( $\xi^s(r)$ ),
$$
\xi_{\alpha \beta} = \xi^s(\vert \br_\alpha - \br_\beta \vert )
= { 1 \over (2 \pi)^3 } \int P(k) \exp \bigl( -(k R_L)^2 \bigr)
\exp \bigl( {\rm i} \bk \cdot (\br_\alpha - \br_\beta ) \bigr) \rd^3k , \eqno(7)
$$
and the shot noise covariance matrix is:
$$
\sigma_{\alpha \beta} = {1\over \bar n (2 \pi R_L^2)^{3} }
\int { 1\over \phi(x) }
\exp \bigl(- { (\br_\alpha-\bx )^2 + (\br_\beta-\bx )^2
\over
2 R{^2_L} }  \bigr)   \rd^3x     \eqno(8)
$$
(The derivation of the error matrix follows Scherrer and Bertschinger 1991.)
Note that the   kernel introduces off-diagonal terms in the error covariance
matrix (ZHFL). The cross-correlation of the high resolution    field and the low
resolution data   is:
$$
\xi_\alpha(\br_i) = { 1 \over (2 \pi)^3 } \int P(k)
\exp \bigl( - { (k R_S)^2 + (k R_L)^2 \over 2 } \bigr)
       \exp \bigl( {\rm i} \bk \cdot (\br_i - \br_\alpha ) \bigr) \rd^3k
\eqno(9)
$$
Defining the WF operator $W_{i\beta}$,
$$
W_{i\beta} =  \xi_\alpha(\br_i) \Bigr(\xi_{\alpha \beta} +
 \sigma_{\alpha \beta} \Bigr)^{-1},  \eqno(10)
$$
a linear high resolution realization is now obtained by
$$
\delta(\br_i) = \tilde \delta(\br_i) + W_{i\beta}
\Bigl( \Delta_\beta - \tilde\Delta_\beta \Bigr).  \eqno(11)
$$

\bigskip

{\bf c. Quasi-Linear Approximation}

\medskip

In the bottom-up model of gravitational clustering small scales go non-linear before
large ones. Hence, our basic approach here is to smooth the data on a scale $R_L$ which
is large enough that the $\delta$ field, smoothed on that scale, is approximately
linear. Given an estimator of such a linear observable, the linear formalism of \S IIa
can be used to make high resolution CRs, as if the linear theory is valid on these
small scales. However, as it will be shown below, numerical N-body  simulations show
that even for quite high $R_L$ smoothing, the resulting field has undergone some non-linear
evolution. Thus, the linear procedure of the WF/CR has to be supplemented by an
additional step of 'linearizing' the input data, \ie\  mapping the present epoch data
back to the linear regime.
Various
algorithms have been proposed to trace back non-linear perturbation field to the linear
regime (\cf\  Strauss and Willick 1995). All of these  `time machines' recover the
initial linear field in the case where the quasi-linear field is known {\bf exactly},
with no statistical uncertainty. The case of real observational data   where the shot
noise
 increases with distance, poses a much more difficult problem. As the density
field is sampled
further away it becomes more dominated by the shot noise and in the mean its amplitude
increases with distance. Thus, a procedure has to be developed that accounts for the
statistical noise, separately from the non-linear effects.

The analysis of the various non-linear effects and the test of possible approximations to
the mapping from the non-linear to the linear regime  is best done by  an  analysis of N-body
simulations of non-linear gravitational clustering. As our aim here is to apply the
method to the \iras\ 1.2Jy
catalog,  N-body simulations with the \iras\  power spectrum     have
been performed. Mock \iras\ data sets are generated  from these simulations.
(Full description of the simulations is given in \S III.) The non-linear evolution has been studied by smoothing the fully
non-linear field, $\delta^{NL}$, and the linear field,
$\delta^L$, on the $R_L$ scale. An examination of the   ($ \delta^{NL},
\delta^L $)  relation (Fig. 5a) shows both a
scatter and systematic deviation from the desired $\delta^L = \delta^{NL}$.
The  empirical fix to the `non-linearity' of the smoothed data
which is used here consists of two steps. First, to account for the scatter in
the ($ \delta^{NL}, \delta^L $) relation   a
new term is introduced to the data auto-covariance matrix, $\sigma^{NL}$. Dealing
with the scatter by statistical means is a manifestation of our inability to invert the
exact non-local non-linear mapping from the linear to the quasi-linear regime.
Here we go to the extreme simplification and take
$\sigma^{NL}  $ to be a scalar   matrix. The value of this constant
term is found to be of the order of $10^{-2}$ and it is determined by the
requirement that $\chi^2/d.o.f. = 1$, where
$\chi^2 =  \vd^\dagger\ \Bigl<\vd\,\vd^\dagger \Bigr>^{-1} \bd \  $
takes into account the theoretical variance, shot noise and  $\sigma^{NL}$.
A WF estimator
of the $R_L$-smoothed field is obtained by applying a WF on the data, where $R_S$ is
replaced by $R_L$ to obtain low resolution,
$$
\delta^{\rm WF,QL}(\br_i)=  \Biggl[ W_{i\alpha} \Biggr]_{R_S=R_L} \Delta_\alpha.
\eqno(12)
$$
The estimation of the quasi-linear correction is given by
$\bigl(\delta^{\rm WF,QL} - f\left(\delta^{\rm WF,QL}\bigr)\right)$ where
$f(\delta)$ is a polynomial fitting to the curve shown in Fig. 5a.
This correction is evaluated at grid points $\br_\alpha$ and is used to correct
the data points:
$$
\Delta{^L_\alpha}=  \Delta_\alpha -
\bigl(\delta^{\rm WF,QL} - f\left(\delta^{\rm WF,QL}\bigr)\right).
\eqno(13)
$$
The modified (`linearized') $\Delta{^L_\alpha}$'s are now substituted in Eq.  11 to
obtain a high resolution CR of the underlying linear field, given the actual data.

The {\it ad hoc} linearization procedure presented here behaves correctly
in the following limits. In the case of distant data points, where the data is dominated by shot noise, the WF attenuates the estimated
field towards zero amplitude. The linearization opperation (Eq. 13) would
hardly change the amplitude of the WF estimator (Eq.12). The WF/CR is therefore dominated
by the  random residual, and consequently the resulting realization lies in the linear
regime. For nearby data points  where the shot noise is negligible,  the WF leaves the
signal almost untouched $ \delta^{\rm WF,QL} \approx \Delta $. The procedure described
here   correctly recovers the Gaussian 1-point distribution function of the density field,
however it will not correctly reconstruct the two point function. The procedure suggested
here works only for data close to the linear regime.

The  mock data base of Eq. 4  that is used to
sample the residual from the mean, has to be subjected to the same non-linear effects as
the  real data do. Thus, the mock data $\tilde \Delta_\alpha$ have to be sampled from a
mock \iras\ catalog drawn out of an N-body simulation, and then to be mapped to the linear
regime in the same way the actual data are treated. The linear CRs  thus generated are used
to set the initial conditions for N-body simulations.

\bigskip
\cntl{\bf III. Testing Against N-body Simulations}
\bigskip

Within the linear regime the WF/CR method provides a rigorous way of estimating the
underlying field and making realizations that are consistent both with the data and the
\prior\ model.  However, beyond the linear regime one should use
various approximations to achieve such a goal. In particular, the mapping of the
quasi-linear data to the linear regime has to be calibrated by testing it against
N-body simulations. This empirical approach strongly depends on the nature of
the data and the assumed model, and therefore it should be tested on mock data sets
that are drawn from the full N-body systems in a manner that mimics the actual
observations. It follows that unlike the linear regime reconstruction, where a general
WF/CR formalism can be formulated, in the quasi-linear domain the approach should be
fitted to the problem at hand.

Our aim is to apply the present method to the \iras\ 1.2 Jy redshift survey. The \prior\
model assumed here is a flat, $\Omega_0=1$, CDM-like model with a shape parameter of
$\Gamma=0.2$,  normalized by $\sigma_8 = 0.7$ and with no biasing. Four random linear
realizations of this model were generated and used as initial conditions to a
$128^3$   PM N-body code (written by E. Bertschinger) with a
comoving box size of $1.6 \times 10^4 \, km/s$, periodic boundary conditions
and grid spacing of $125 \, km/sec$. The low
resolution is defined by a Gaussian filter of smoothing length
$R_L=1000 \, km/s$ (here
distances are given in units of $\, km/s$)and the high resolution is limited by
the  Nyquist wavelength.
The   smoothed field is sampled within a sphere of radius $6000 \, km/s$ on a
Cartesian grid at a sampling rate of $R_L$, yielding   $925$ constraints. The
smoothing is done by an FFT convolution on a count-in cells (CIC) of the
particles done on the basic grid.   In the numerical experiments no Galactic
ZOA is assumed.

A non-trivial task here is the
calculation of the non-diagonal error correlation matrix. The symmetry of the
problem leaves only three independent degrees of freedom out of the six
(${\bf r}_1, {\bf r}_2$) variables. The resulting integral   is taken over  a
product of a  Gaussian that peaks at  $\sim ({\bf
r}_1 + {\bf r}_2)/2$   and the inverse of the selection function, which
makes it a three-dimensional integral. The integral is calculated numerically
on a finite cube of size $ 6R_L$ centered at $\sim ({\bf r}_1 + {\bf r}_2)/2$.
Special care has been taken to control the numerical errors, as the inverse of
the auto-correlation matrix is found to be  sensitive to the numerical noise.
The matrix inversion is done by a  Cholesky decomposition algorithm which is fast
and stable (Press {\it et al.} 1992).

The
numerical simulations provide  an ensemble of four random non-linear
realizations of
the \prior\ model. The first step in analyzing the N-body simulation is the
construction of the `galaxy' distribution out of the `dark matter' particles.
No biasing is assumed here and a fraction of the particles are randomly tagged
as galaxies, so at to reproduce the \iras\ mean number of galaxies. Here a mean
number density of $ n_g=5.48 \times {10^{-8}}
{{(\, km/s)^{-3}}}$ (Fisher 1992, Fisher {\it et al.} 1994) is used.
Given a `volume limited' mock \iras\ catalog, a selection process is
applied to generate a magnitude limited mock \iras\ 1.2Jy catalog, based on the
selection function:
$$
\phi(r)= {{\left(r\over{r_s}\right)}^{-2\alpha}} \,
{{\left({{{r^2}+{{r_*}^2}}\over{{{r_s}^2} + {r_*}^2}}\right)}^{-\beta}}    \eqno(14)
$$
with ${r_s} = 500 \, km/s, \alpha=0.492, \beta=1.830, {r_*} =
5184 \, km/s$ (Yahil {\it et al.} 1991, Fisher 1992).

\myfig {1} {14.0}
{\includegraphics{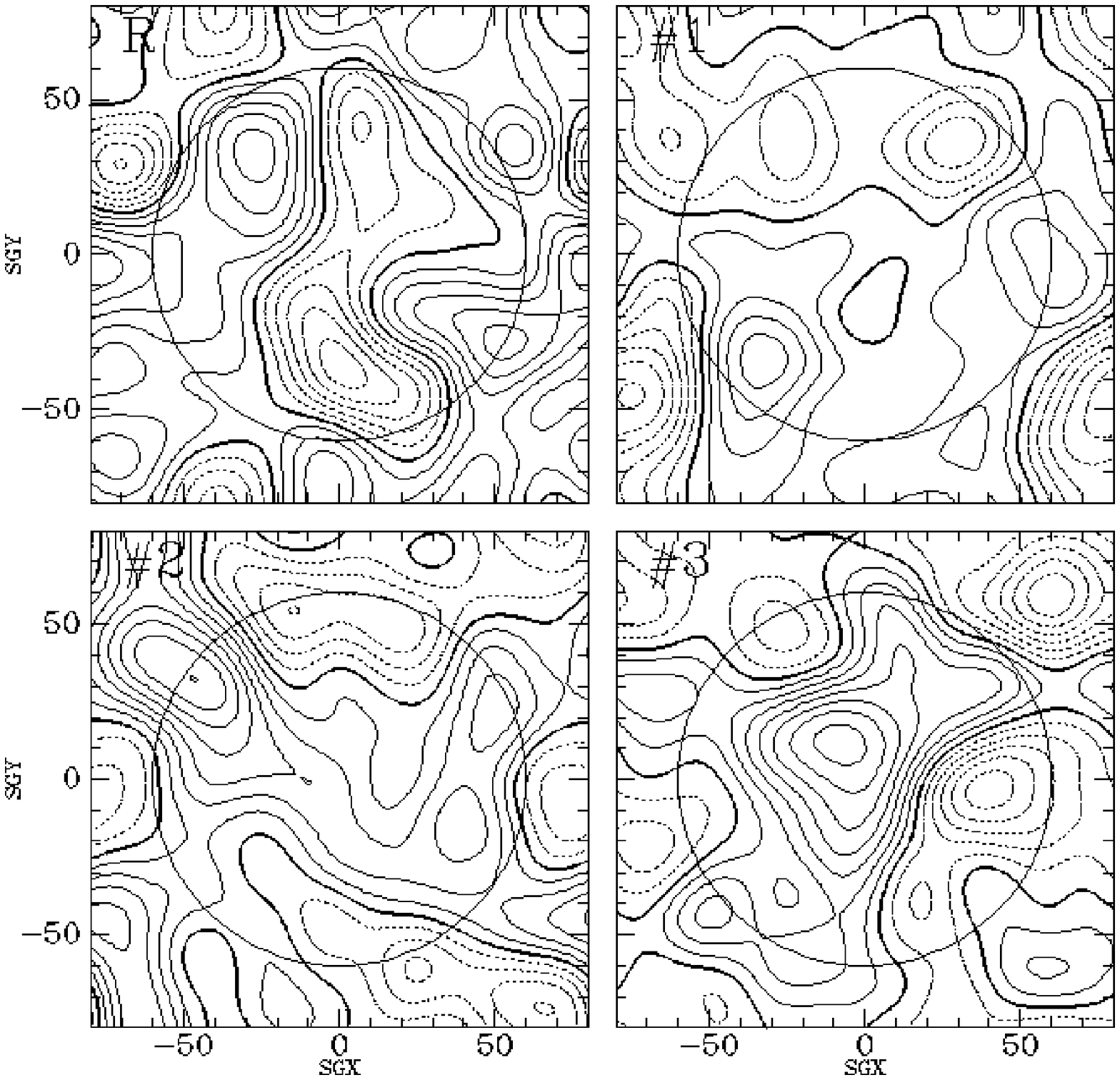}} {-2.0}
{Random fields: The linear fluctuation density fields of the four
random realizations given the {\it prior} as described in the text, smoothed
with a $1000 \, km/s$ Gaussian window. The Supergalactic plane is shown, the units
are in $100 \, km/s$ and the contour spacing is 0.2 in $\delta$. The heavy contour
belongs to $\delta=0$ while solid and dot contours represent overdensity and
underdensity regions respectively. The circle of radius $6000 \, km/s$ marks the
region from which constraints are taken.}

From the four different
realizations one is treated as the `real' data and other three are used to make the
NLCRs. These are labeled by `R' and 1, 2 and 3 in the various figures. The four
different linear random realizations are shown in Fig. 1. (In all figures the
continuous $\delta$
field is shown at the low resolution of $R_L=1000 \, km/s$ with contour spacing
of 0.2 in $\delta$). The `R' non-linear realization
has been sampled to produce a mock \iras\ data base, from which the CRs are generated.
The
three linear CRs, which serve as initial conditions to the PM code, are shown in Fig. 2
where the R-realization is given as well.
The success of the whole procedure is
actually determined at this stage. Namely how close are the various 1, 2, 3
fields to the 'R' field. The  general `cosmography' is nicely
recovered by the different CRs, with a little scatter. The smoothed density of the fully evolved NLCRs and the original 'R' field are shown in Fig. 3.  Fig. 4 shows the `R' volume
limited non-linear `galaxy' distribution, which is to be recovered by the three NLCRs. The
bottom panel shows the magnitude  limited `galaxy' distribution, using the \iras\
selection function, from which the constraints are drawn. Various objects that dominate
the `cosmography' of the simulations are tagged by capital letters A-F in
Figs. 2, 3 \& 4. The circle of radius $6000  \, km/s$ marks the region from which
constraints are drawn.

\myfig {2} {14.0}
{\includegraphics{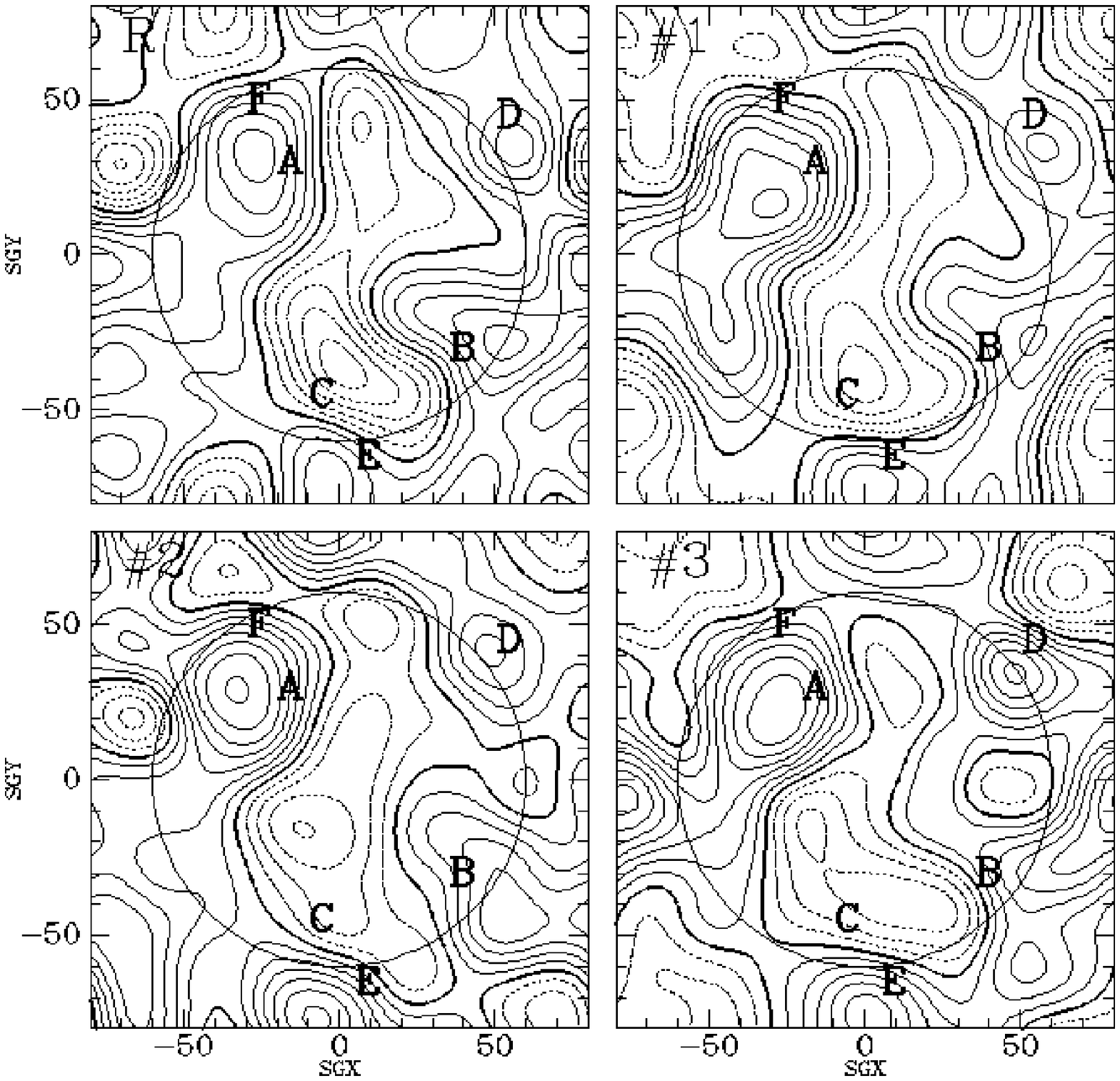}} {-2.0}
{Linear CRs from non-linear data and the linear `R' field: The linear CRs (1,
2, 3) are constrained  by data sampled from the    fully evolved 'R' field. These are
supposed to recover the linear `R' field, which is presented here. The various objects
that define the gross structure are marked by    the letters A, B, C, D, E \& F, and these
are introduced for the sake of the  comparison of the reconstructed fields with the
original. The parameters of the plot are the same as in fig. 1.}

The quality of the reconstruction provided by the NLCRs can be judged from Figs. 3 \&
4. The main objects that define the cosmography of the 'R' simulation, within the
sampling radius of $ \approx 6000 \, km/s$  , are the A, B, D \& E clusters
and the extended void that occupies most of the sampling volume (C). The F denotes a
filament running from A towards the object at $SGX, SGY=(0,8000)$, which is clearly seen in
the particle distribution but is hardly noticed in the smoothed field.  At the level of
the smoothed field, the NLCRs (1-3) recovers the R realization to within one or two
contour lines.  No systematic discrepancy between the recovered fields and the original
one is found here.  The comparison at the particle
distribution level constitutes a much more challenging test to our method. The robust
peak at A breaks into  sub-clumps that are located at the intersection of a few
filaments. The actual sub-clumps position and distribution vary in the NLCRs, but the
overall statistical nature of the objects is recovered. As expected, the reconstruction
of the filaments is less robust than the density peaks. The F filament that starts at
one side of the A cluster and continues at its other side is reproduced by all the
NLCRs. This is also the case with the filament that runs from cluster B towards the
origin. However, there is quite a large scatter in the overall structure of the network
of filaments.  The conclusion that follows   is that the NLCRs
provide a faithful reconstruction of the non-linear density field at the $1000   \, km/s$
Gaussian smoothing level. The highly non-linear small scale structure is formed within
the correct `boundary conditions' provided by the `actual universe' (\ie\ the 'R'
realization), but with some degree of random variability as dictated by the
unconstrained random realizations used.

\myfig {3} {14.0}
{\includegraphics{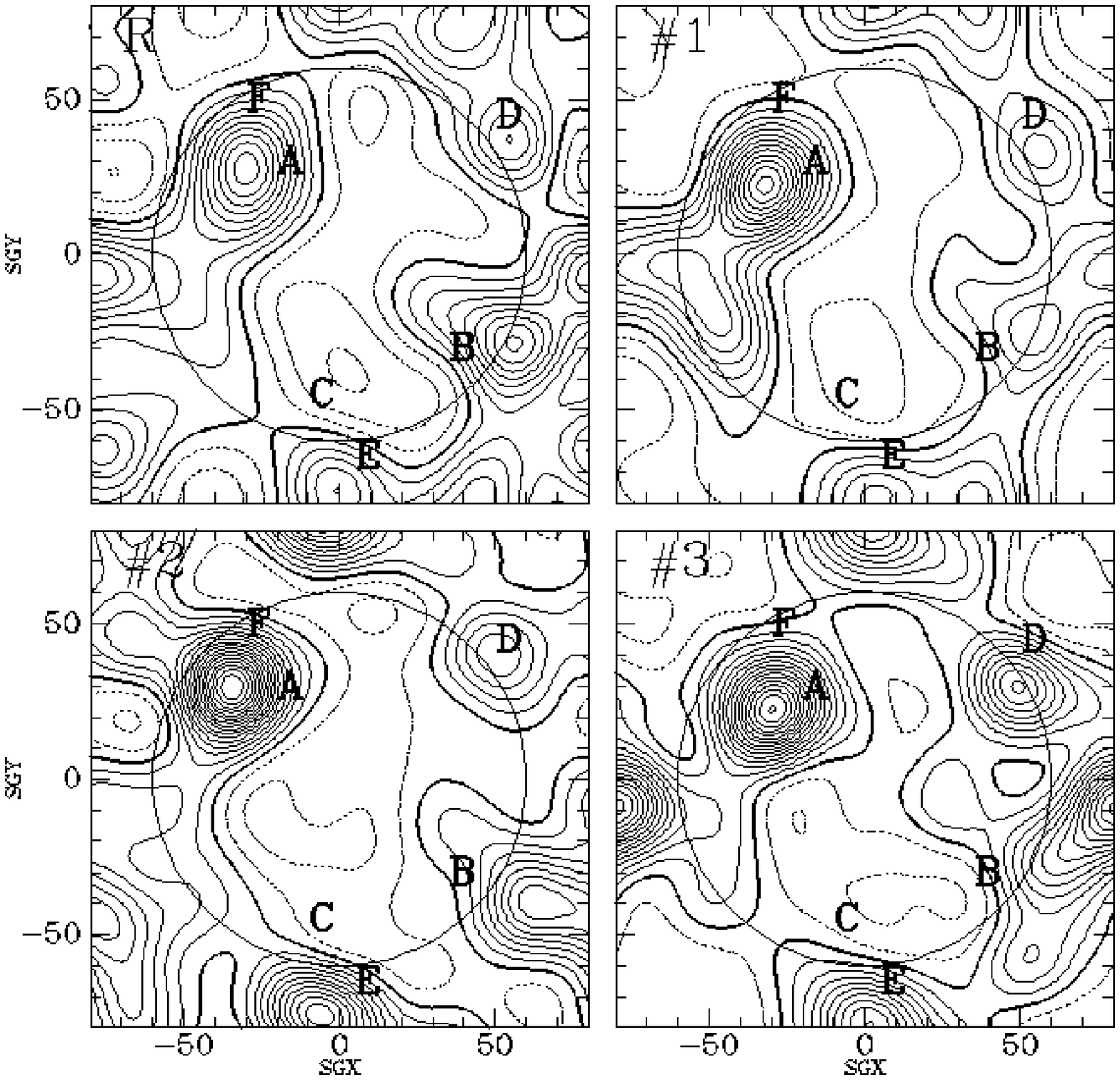}} {-2.0}
{NLCRs and the non-linear `R' field: The smoothed non-linear evolved density
fields of the {\it real} universe, 'R', and the   ensemble of NLCRs(1,2 \& 3). The NLCRs
reconstruct the `R' field, within the limitations of the method.}

\myfig {4} {14.0}
{\includegraphics{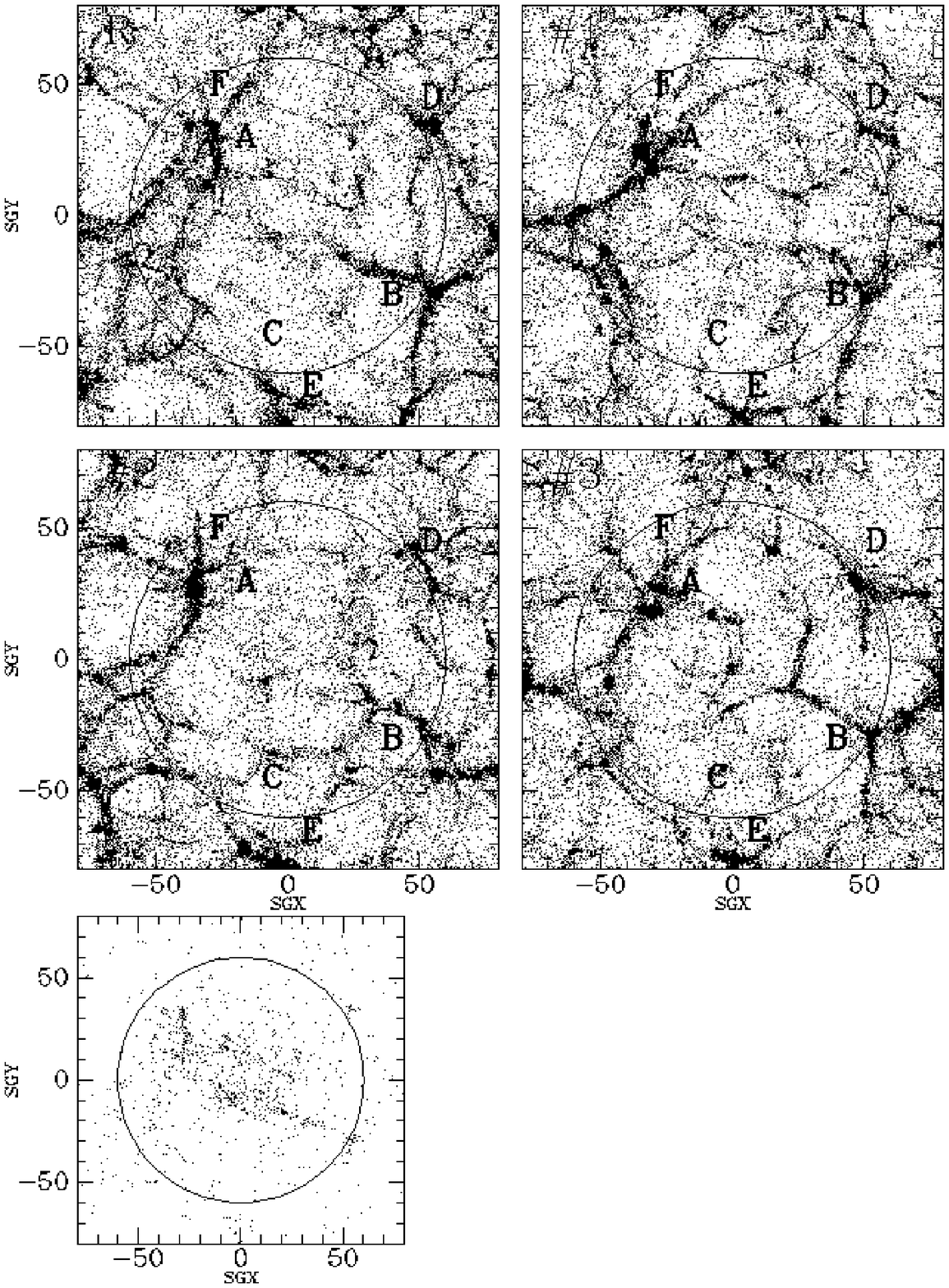}} {0.0}
{Volume limited mock `galaxy' catalogs: Slices of thickness of $2000 \, km/s$ of
the volume limited 'galaxy' distribution normalized to the mean galaxy density
of  the \iras\ 1.2Jy survey. The top left panel shows the `galaxy' distribution of the
`R' field. Applying the \iras\ selection function produces the magnitude limited sample
that is shown in the bottom left panel. The mock survey is used to set constraints on the
random realizations of the underlying field. The `galaxy' distributions of the three NLCRs
are shown here, and these are to be compared with the `R' field.}

The simulations are used to calibrate the non-linear correction of Eq.13.
 A scatter plot of the $(\delta^L , \delta^{NL})$ is shown in Fig. 5a (left
panel), where both the bias and scatter are clearly seen. The non-linear mapping is
applied to $\delta^{NL}$ to produce a corrected linear $\delta^{CL}$. Indeed, the
scatter plot of  $(\delta^L , \delta^{CL})$ (Fig. 5a, right panel) shows that the bias
is removed without increasing the scatter (see also Nusser \etal\
1991). The mapping recovers the normal 1-point
distribution function (Fig. 5b) and the theoretical power spectrum (Fig. 5c).
Here only the dynamical aspects are studied, and no `observational' uncertainties are
considered.

\myfig {5} {15.0}
{\includegraphics{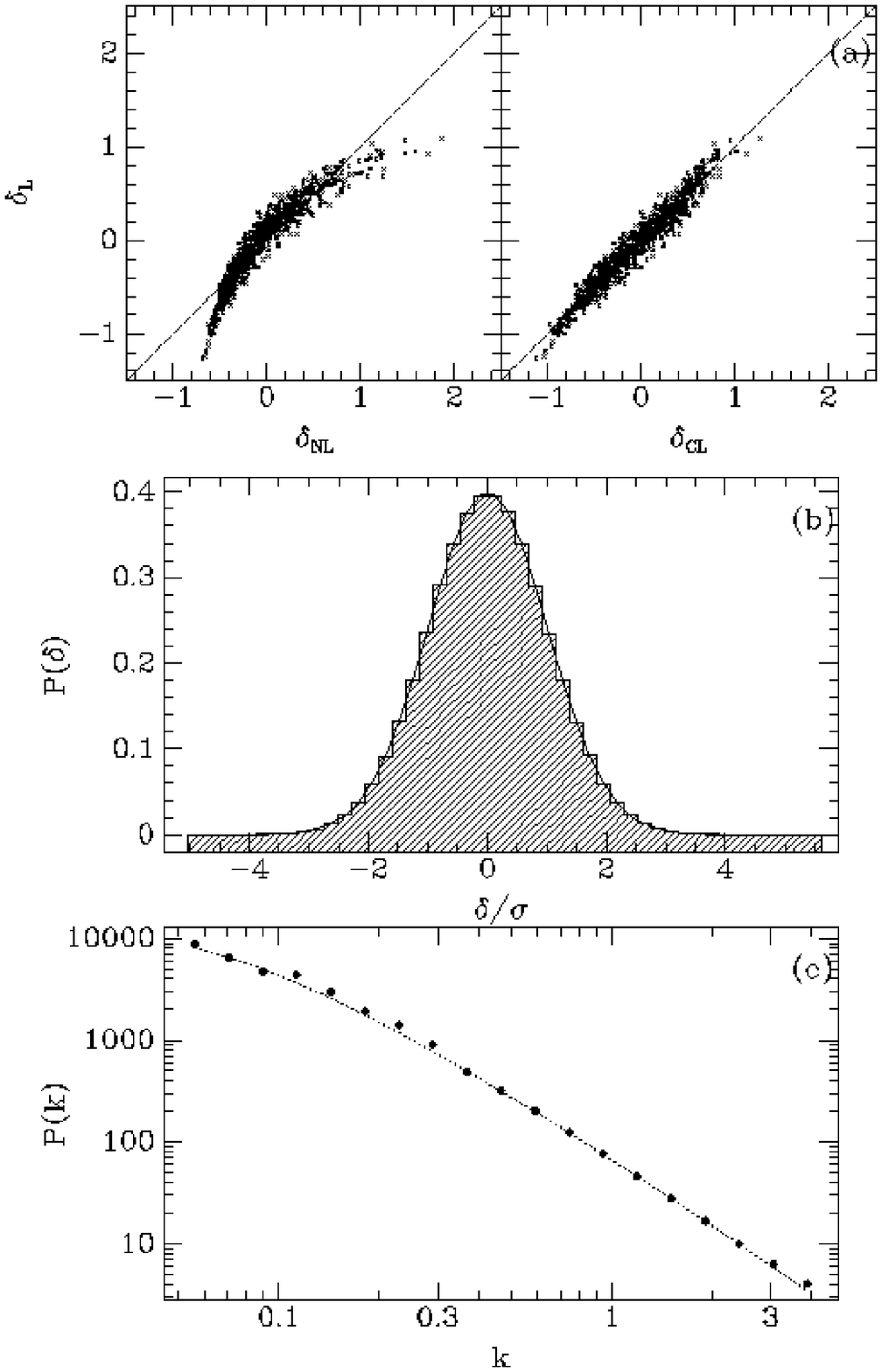}} {0.0}
{(a) {\it Left:} Scatter plot of linear vs. non-linear evolved
density field smoothed with a $1000 \, km/s$ Gaussian filter. {\it Right:} The
same plot after the application of a polynomial correction on the non-linear
field. (b) The 1-point probability distribution function of $\delta$
(histogram) as calculated from the linear CR given corrected linear
constraints. The solid line represents the theoretical Gaussian PDF. (c) The
power spectrum of the linear CR (points). The dotted curve is the CDM-like power spectrum that is used as a \prior\ for the reconstructions.}

The mock realization has been used to simulate the effect of the improvement of the
sampling on the quality of the reconstruction. The 'R' simulation has been re-sampled by
the \pscz\ selection function, which corresponds to an \iras\ survey sampled down to a
0.6Jy flux at $60\mu$. (The \pscz\ survey is an extension of the  0.6Jy  QDOT 1-in-6 survey,
 Saunders {\it et al.} 1990 , which is
now being completed  to all galaxies brighter than the above limit. For details
see {\tt  http://www-astro.physics.ox.ac.uk/}${\sim}${\tt wjs/pscz.html}.)
The improvement in the errors is shown in Fig. 6 as they are normalized to the
{\it rms} value of the underlying field.
Here, only one NLCR has been reconstructed, and the No.1 realization is used. Fig. 7
shows the smoothed 'R' realizations, to be reconstructed, the linear CR that serves as
the initial conditions, the smoothed density field and particle distribution of the
NLCR.
A comparison of Fig. 7 with the Fig. 3 \& 4 shows the   improvement gained by the
reduction of the shot-noise. This is clearly seen in the smoothed non-linear field, where
the PSCZ-based reconstruction recovers the original field much better than the 1.2
Jy-based field.

\myfig {6} {14.0}
{\includegraphics{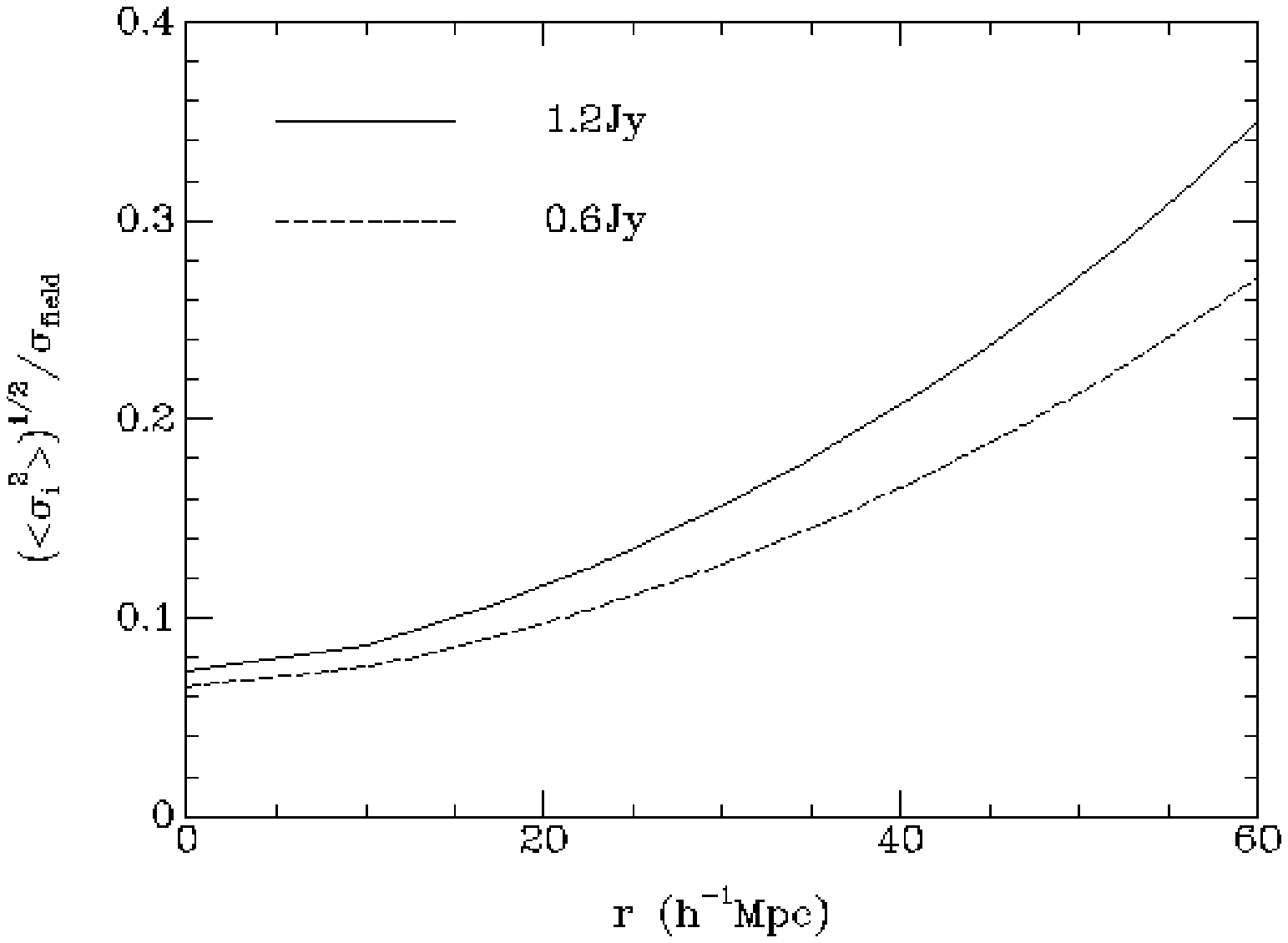}} {-6.5}
{The diagonal elements of the error correlation matrix,
normalized to the {\it rms} value of the $1000 \, km/s$ smoothed field vs. the
distance. The two curves show the variation of the errors in the case of the
\iras\ 1.2Jy and \pscz\ selection functions.}

\myfig {7} {14.0}
{\includegraphics{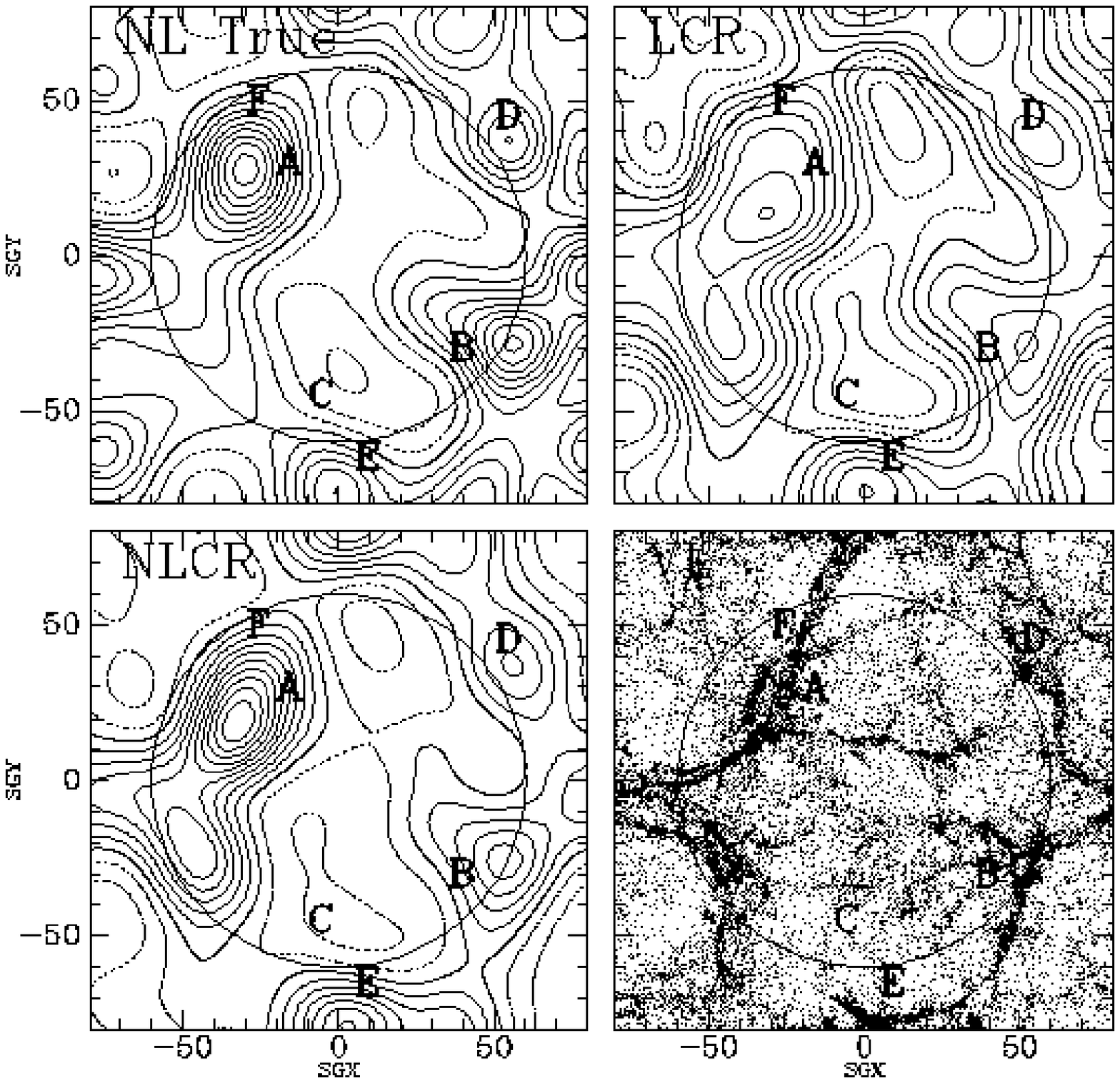}} {-4.0}
{\pscz\ reconstruction of the mock catalog: {\it Upper left:} The
non-linearly evolved `R' field. {\it Upper right:}
Linear CR from a
\pscz\-like survey of the `R' field. The constraints are applied to the random
realization \# 1 of the ensemble of realizations. {\it Lower left:} The smoothed NLCR field
for the \pscz\ case. {\it Lower right:} Slice of thickness $2000 \, km/s$ centered at the
Supergalactic plane of the reconstructed volume limited PSCZ-like survey.}

\bigskip
\cntl{\bf IV. Volume Limited \iras\ Catalog}
\bigskip

The \iras\ 1.2Jy catalog consists of 5321 galaxies. These are used to evaluate the
smoothed density field on a Cartesian grid of $1000 \, km/s$ spacing within a $6000
 \, km/s$, excluding the ZOA, yielding 846 constraints.  The NLCRs are
created on a finer
$128^3$ grid of $125 \, km/s$ spacing, assuming periodic boundary
conditions. However, for a CDM-like power spectrum the
structure within the $6000 \, km/s$ is hardly affected by the periodicity on the $\pm
8000 \, km/s$ box. A word of caution should precede the analysis of the NLCRs. The neglect
of the redshift distortions affects the reconstructions in two ways. One is the
displacements of the objects by a few $\times 100 \, km/s$, and the other is the
amplification of the (over)density amplitude, due to the gravitational
focusing (\cf\ Kaiser 1986).

\myfig {8} {14.0}
{\includegraphics{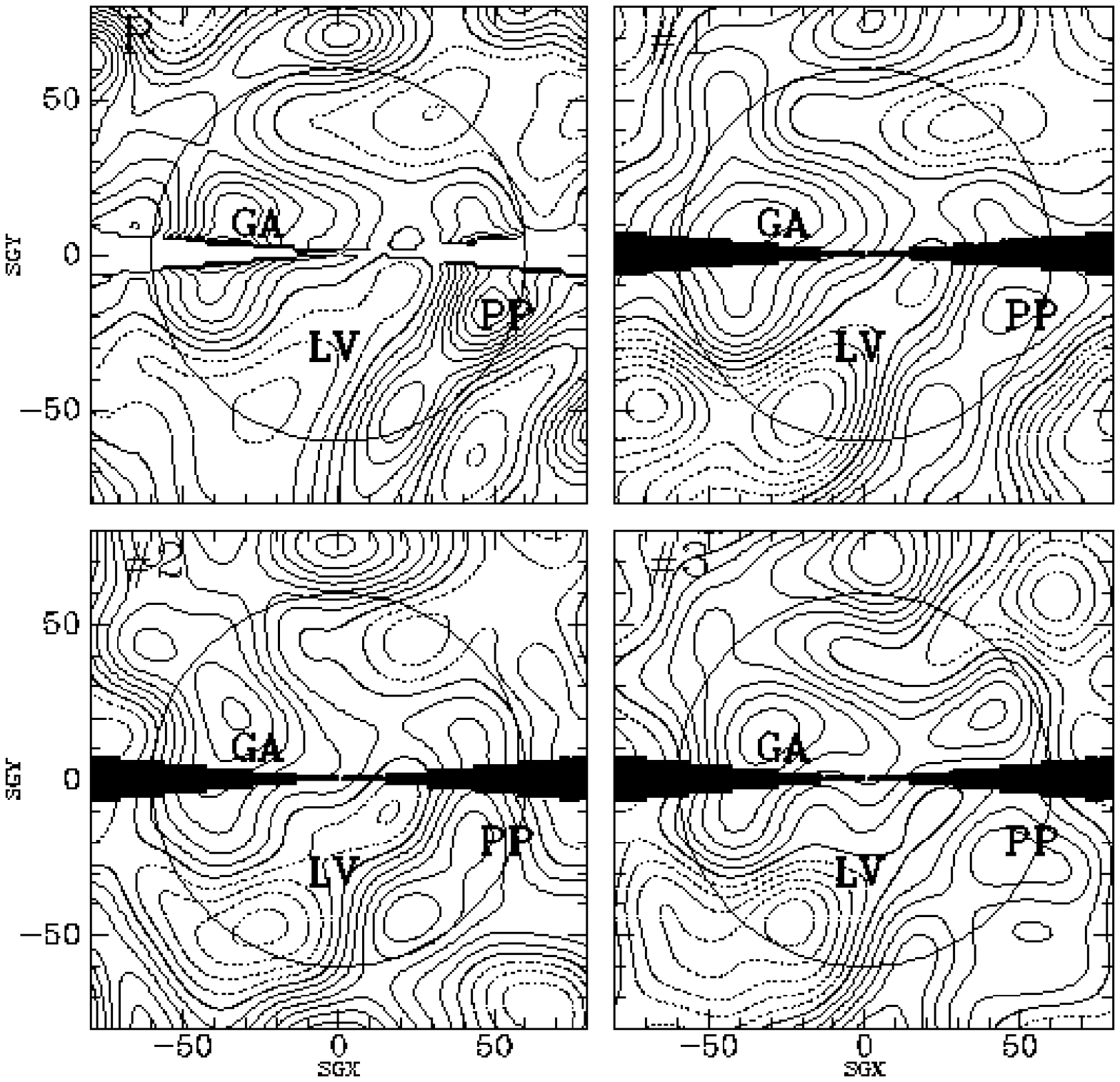}} {-4.0}
{Linear CRs of the \iras\ 1.2Jy survey: The 'R' plot shows the
Supergalactic plane of the $1000 \, km/s$ smoothed raw \iras\ 1.2Jy survey. The rest
three plots show the linear CRs which are reconstructed by imposing  constraints from
the smoothed actual \iras\ survey. The three linear CRs are generated from the three
random realizations (1,2, \& 3) used for the reconstruction of the mock catalog.}

The actual \iras\ survey is affected by Galactic ZOA  of  $\vert b
\vert \le  5^\circ $. Our choice of a Cartesian three dimensional representation is
especially
suited for handling zones of missing data, where the only correction to be applied is for
the truncated Gaussian smoothing   near the ZOA. Otherwise,
no assumption is made on the completeness of the survey sky coverage. This is to be
contrasted with the case of  orthogonal representations such as
Fourier or spherical harmonics/Bessel functions, where the treatment of regions of missing
data usually results  in a very complicated mode-mode coupling  (Fisher \etal\ 1995a)
The Wiener filter extrapolates the density field within the ZOA, using the galaxy
distribution at the two sides of the ZOA.

\myfig {9} {14.0}
{\includegraphics{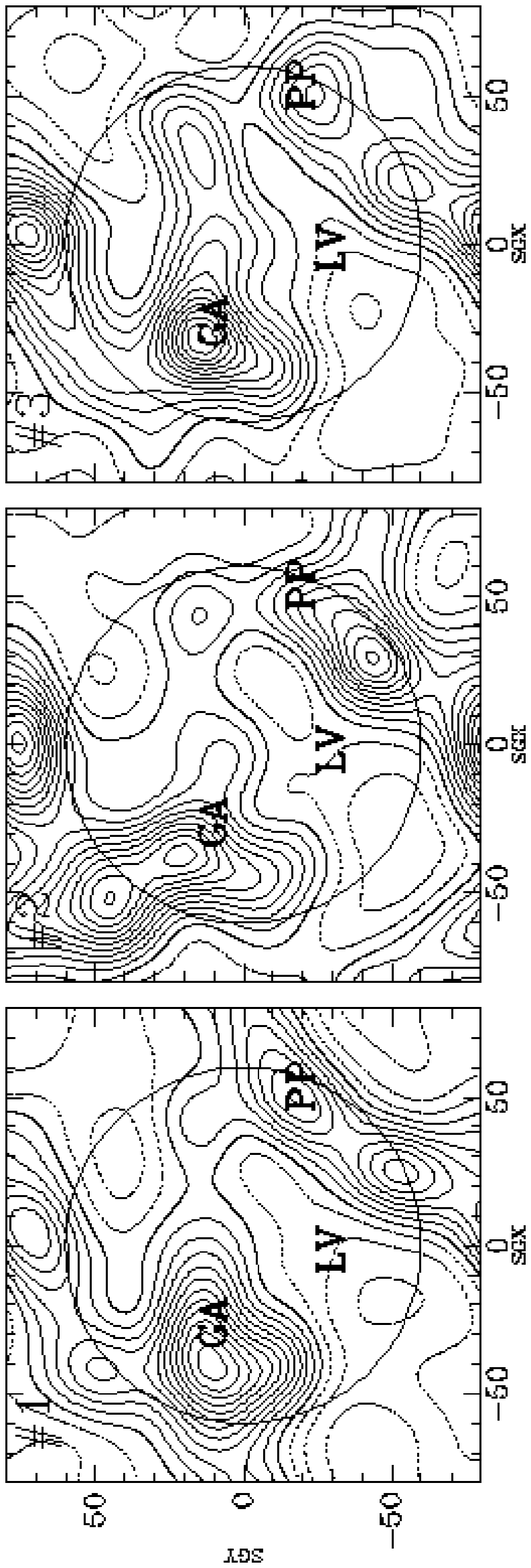}} {-8.0}
{NLCRs of \iras\ 1.2Jy: The ensemble of the three NLCRs based on
the \iras\ 1.2Jy survey smoothed by a $1000 \, km/s$ Gaussian window.}

An `ensemble' (namely, three) of NLCRs has been generated, from which one can estimate the
variance of the different reconstructions and thereby asses the quality of the
reconstruction. The `observed' smoothed \iras\ 1.2Jy density field is presented in Fig. 8
(panel labeled by R), and the three  linear CRs are shown as well, all presented on the
Supergalactic plane. Note that in the linear CRs   the amplitude of the
positive (over) density field never exceeds the observed one. The voids, on the other
hand, become more empty in the CRs because of the non-linear correction. The three NLCRs
of the actual universe are shown in Fig. 9. The random realizations used here are the same
ones used for the reconstruction of the mock data (Fig. 1).  The fundamental features of
the Supergalactic plane are  the Great Attractor (GA), the Perseus part of the
Perseus-Pisces (PP) supercluster, and the Local Void (LV).
The overdensity containing the
Coma cluster at $SGX,SGY \approx (0,8000) \, km/s $ is clearly seen. However, it lies at the
edge of the simulation and is   affected by the periodic boundary conditions and therefore
its physical characteristics cannot be studied here. The   `volume limited' galaxy
distribution of the NLCRs as well as the \iras\ 1.2Jy   survey itself are shown in Fig.
10. The   structure of peaks and the network of filaments at the GA region, on the one
side of the   Local Group  and at the PP region  on the other side, is a robust
feature of all the NLCRs. The local void is recovered in all the NLCRs, but the details of the
filamentary network within the voids vary for one realization to the other.
It is in  this sense that    the term of `non-linear reconstruction'
of the local LSS can be used, namely the gross features of the LSS are imprinted onto the
otherwise random non-linear realizations.

\myfig {10} {14.0}
{\includegraphics{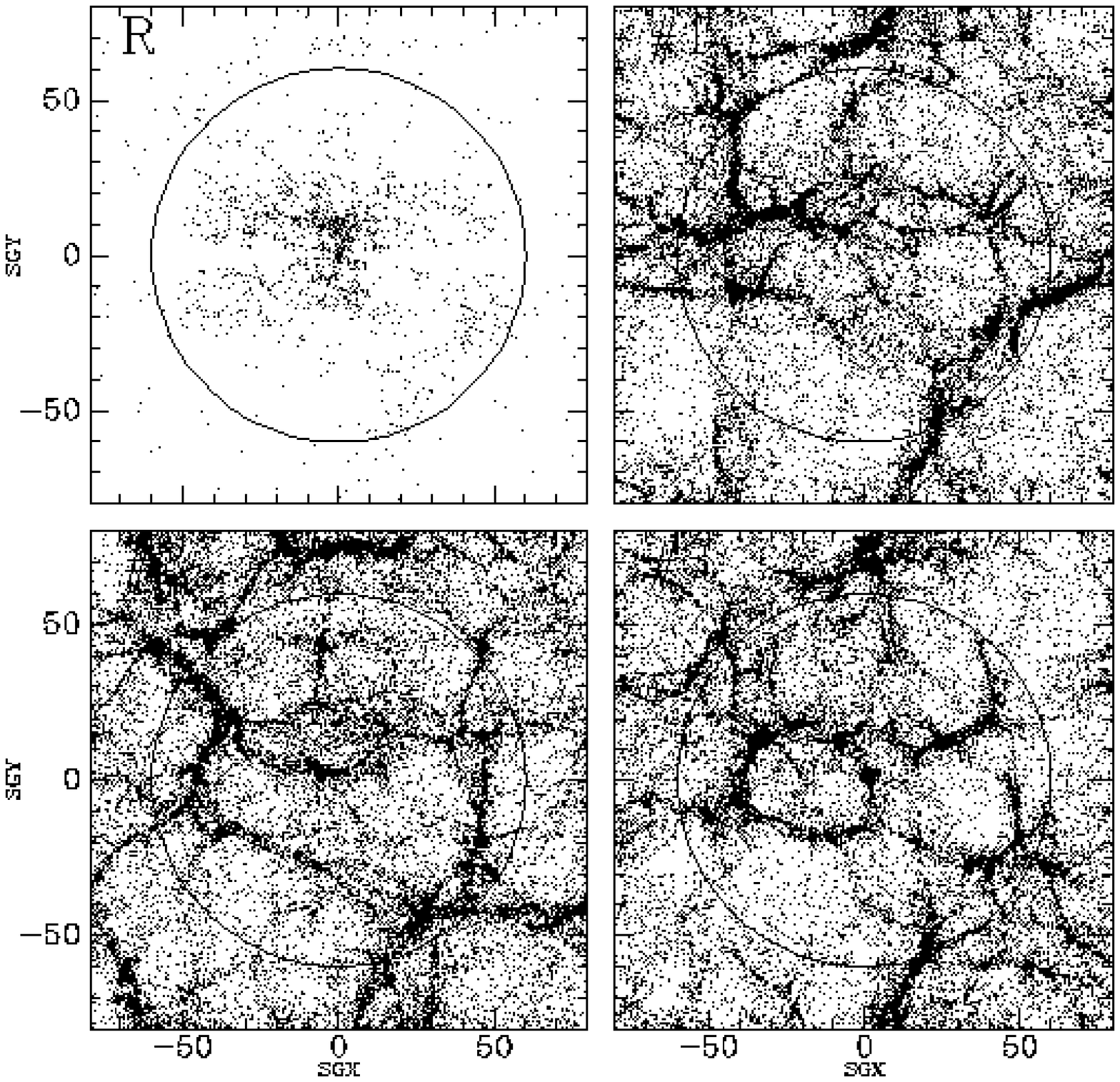}} {-4.0}
{Volume limited \iras\ 1.2Jy reconstructions: The 'R' plot shows a $2000
\, km/s$ thick slice of the \iras\ 1.2Jy redshift survey centered at (SGZ=0). The other
plots show the ensemble of the reconstructed volume limited \iras\ 1.2Jy surveys.}

To further study the cosmography recovered by the NLCRs, Aitoff projections of the
`galaxy' distribution are plotted. Here only one NLCR (No.1) is analyzed in that way.
Aitoff projections of the `galaxy distribution' within  shells of thickness of $1000 \, km/s$ at distances of $2000, 3000$ and $4000 \, km/s$ (Fig. 11a) and $5000, 6000$
and
$7000 \, km/s$ (Fig. 11b) are plotted. Contours of the smoothed field are superimposed
on the `galaxy' distribution. Variable smoothing is used here   to compensate for
the variation of the angular particle density with  depth. The smoothing kernel radii
used here  are $500, 640, 780, 920, 1060, $ $1200 \, km/s$, corresponding to
$\delta_{rms}=0.80, 0.67, 0.56, 0.46, 0.41, 0.37$. In all projections a contour spacing
of $0.2$ is used. The identification of the various objects grossly follows the
cosmographical analysis of Webster, Lahav and Fisher (1996), who used a  linear Wiener
filter in
 the spherical harmonics/Bessel representation to analyze the same data base used here.

\myfig {11a} {14.0}
{\includegraphics{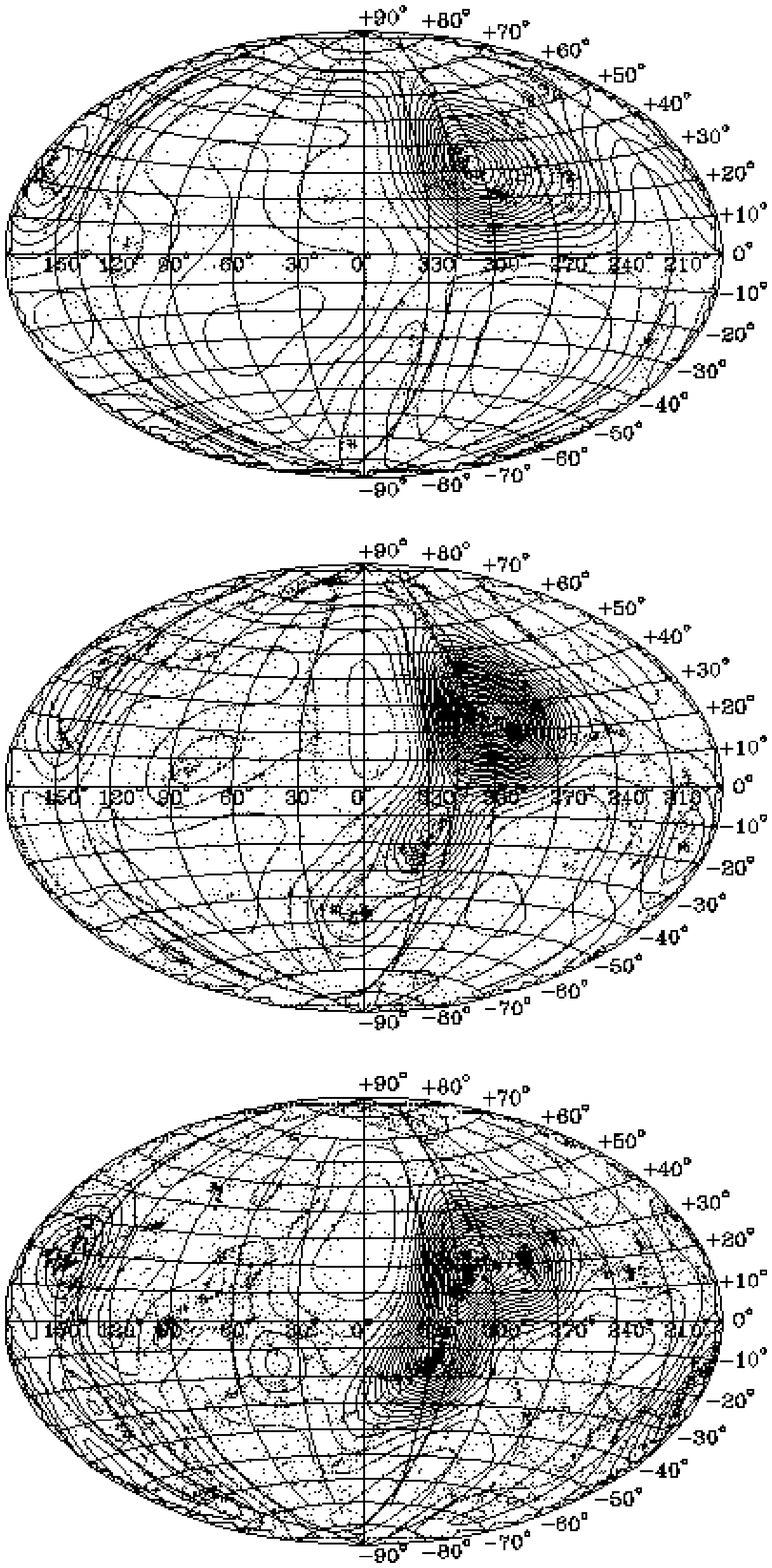}} {0.0}
{Aitoff projections in galactic coordinates of the
 NLCR (\# 1)
reconstruction of the volume limited \iras\ 1.2Jy: From top to bottom the $2000,
3000$   \& $ 4000\ \, km/s$ shells. The
thick curve marks the Supergalactic plane and the particles are within a $1000 \, km/s$ thick
slice centered on these distances.}

The nearby structure, at $2000 \, km/s$, is dominated by the foregrounds of
Centaurus  at $(l,b) \approx (305,30)^\circ $. The
$\ 2000 \, km/s$ plot shows also the Ursa Major $(l,b) \approx (170,25)^\circ $ and the
outskirts of the Fornax-Doradus-Eridanus complex $(l,b) \approx (180,-60)^\circ $.
Underdense regions are identified at the two sides of the Supergalactic plane.
These include the
Local Void   ($330^\circ<l<120^\circ , -70^\circ <b< 60^\circ  $),
a void at $(120^\circ<l<200^\circ,-40^\circ<b<0^\circ)$ and, a void at
$(210^\circ <l< 330^\circ, -85^\circ  <b< 0^\circ )$, which extends over all
the slices  out to $7000 \, km/s$.

The structure at $3000 \, km/s$ is dominated by the GA complex which lies close to
the Supergalactic plane and is almost aligned along the line of sight towards   $(l,b)
\approx  (300,25)^\circ $. The Pavo-Indus-Telescopium (PIT) complex $(l,b) \approx
  (330,-15)^\circ$ forms an extension of the GA. On the other side of the Supergalactic
plane the structure is dominated by   two ridges, one  in the direction of  $(l,b)
\approx (150, 25)^\circ$  connecting  Ursa Major (at $2000 \, km/s$) with
Camelopardalis (at $4000 \, km/s$) and the other that extends all the way towards the PP at
$(l,b) \approx (165,-15)^\circ$, which peaks at a distance of $\approx 6000 \, km/s$.

At $4000 \, km/s$ Centaurus/GA ($\ (l,b) \approx (315, 20)^\circ$) and PIT ($\ (l,b)
\approx (325,-15)^\circ$) are the dominant overdensities,  and these are
connected by a filamentary structure running close to the Supergalactic plane
at $l\sim 320^\circ,\  -10^\circ <b< 10^\circ $. Other features in the plot
are a concentration at $(l,b) \approx (180,-25)^\circ$ and the Camelopardalis cluster
$(l,b) \approx (150,15)^\circ$.

\myfig {11b} {15.5}
{\includegraphics{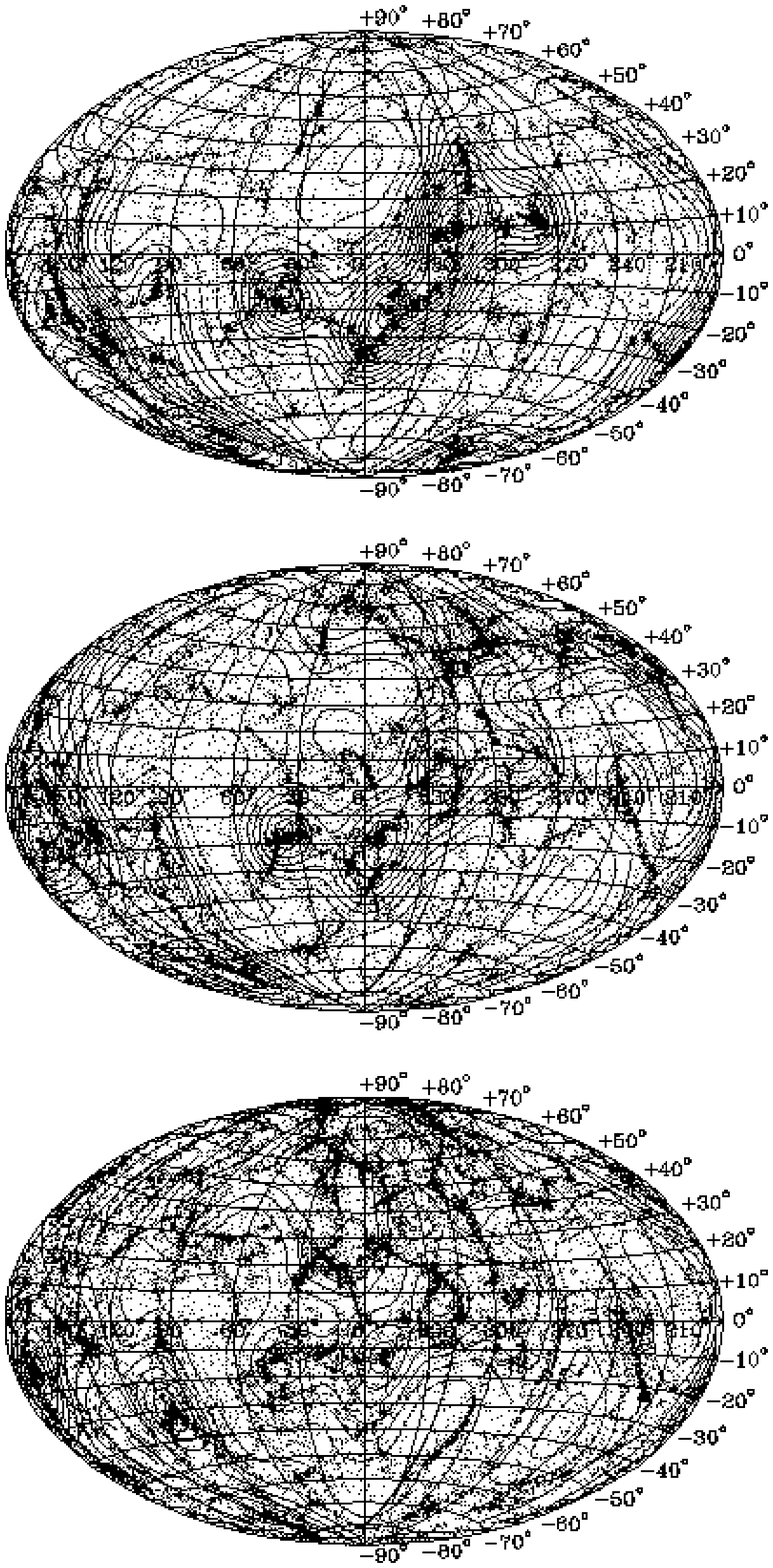}} {0.0}
{Aitoff projections in galactic coordinates of the
 NLCR (\# 1)
reconstruction of the volume limited \iras\ 1.2Jy: From top to bottom the $5000, 6000$
\& $7000 \ \, km/s$ shells. The
thick curve marks the Supergalactic plane and the particles are within a $1000 \, km/s$ thick
slice centered on these distances.}

Going further to $5000 \, km/s$ the plot shows  the Centaurus/GA $(l,b) \approx
(320, 10)^\circ$,   a continuation of the    $(l,b) \approx (180,-25)^\circ$ structure,
the  PP at $(l,b) \approx (150,-15)^\circ$,
a prominent overdensity at $(l,b) \approx (275, 10)^\circ$ which is quite
separated from  Centaurus, PIT $(l,b) \approx (345,-20)^\circ$ continues
from the previous slice, and an overdensity at $(l,b) \approx (130,-65)^\circ$
which  is extending out to $7000 \, km/s$. It is not clear whether  the remarkable
concentration at
$(l,b)  \approx (40,-15)^\circ$, which seems to extend from $4000 \, km/s$ to $7000 \, km/s$
is a real object or a fluke induced by the shot noise.

The \iras\ density field is sampled within  a distance of $6000 \, km/s$ and beyond it the
random nature of the NLCR becomes more dominant.  To study the reconstructed
structure at these distances a  more detailed comparison of the members of the
`ensemble' of NLCRs is needed, so that the robust features of the realizations can
be separated from the random ones. Such an effort is beyond the scope of the present
work. Here only the main objects of the $6000 \, km/s$ slice are noted.
 The
dominant  feature at this shell is the PP at $(l,b) \approx (165,-15)^\circ$. Also
shown in this plot are the 'tail' of Centaurus/GA $(l,b) \approx (320,10)^\circ$,
Leo  $(l,b) \approx (270, 50)^\circ$, Cancer $(l,b) \approx (190, 35)^\circ$
and PIT $(l,b) \approx (350,-15)^\circ$. The strong
overdensity at $(l,b) \approx (40,-20)^\circ$ extends  from the
$5000 \, km/s$ slice to the $7000 \, km/s$ one.

The almost complete sky coverage of the \iras\ survey, of a ZOA of $
\vert b \vert \leq 5 ^\circ $, and the method used here enable a good restoration of
the obscured structure. Indeed the density field found here agrees with optical
studies of the ZOA, in particular the optical survey of the southern ZOA (
$(l,b) \approx (265$ --- $340, -10$ --- $ 10)^\circ$) of Kraan-Korteweg and her colleagues
(\eg\ Kraan-Korteweg, Woudt and Henning 1996).

\bigskip
\cntl{\bf V. Discussion}
\bigskip

The NLCR algorithm presented here enables one to perform controlled Monte Carlo
N-body simulations of  the formation of our `local'   universe. These are designed
  to recover the actual observational LSS  within the
statistical uncertainties of the data. The new ingredient introduced here is the
reconstruction of the non-linear regime, \ie\ the extrapolation in Fourier space from
small to large wavenumbers that are deep in the non-linear regime.

The NLCR introduced here can serve as a tool for studying and analyzing the large scale
structure of the universe. Some of the obvious problems where NLCRs are expected to be
very useful are: (1) The reconstruction of the  velocity field from
redshift catalogs; (2) Mapping the zone of avoidance and extrapolating the dynamical
fields into unobserved regions; (3) Studying the dynamics   of
observed rich clusters with  the actual initial and boundary conditions; (4)
Analysis of filaments and pancakes as probes of   the initial conditions
and the cosmological model; (5) The NLCR can serve as a probe of the biasing mechanism.
The main virtue here lies in the fact that different data sets, which in principal can
represent different biasing of the underlying dynamical field, can be used to
simultaneously set constraints on the realizations.  Given all these and the technical
simplicity of the algorithm we expect it to be a standard tool of N-body and gas
dynamical simulations.

The algorithm used here relies on using a Cartesian spatial representation of the density
field. This should be considered as an attractive alternative to the spherical
harmonics/Bessel functions representation (Fisher \etal\ 1995a). The later provides a
 representation that is well suited to handle full sky redshift surveys, including an
elegant treatment of redshift distortions. However, this method has two main
shortcomings. One is the complicated mode-mode coupling that is introduced by an
incomplete sky coverage. The other is the effective variable smoothing of the
reconstruction, which is not suitable for setting initial conditions for N-body
simulations. These two problems are naturally solved by using the Cartesian
representation, where no {\it a priori} geometry of the survey is assumed, and an
arbitrary sky coverage of various data sets can be handled.  The   representation
used here  provides   CRs   that
have a constant spatial resolution, which makes it the optimal tool for setting initial
conditions for numerical simulations.  Indeed, the amplitude of the mean (WF)  field
decreases as   the shot-noise errors increase with distance, but this is compensated by
the random component of the CRs. The resulting realizations have a constant resolution with
a  constant power. The method outlined here can be extended to handle very large data sets
by using data compression methods, in particular the signal--to--noise eigenmode
expansion (Zaroubi 1995, Vogeley and  Szalay 1996,Tegmark, Taylor, \& Heavens
1996). Using this representation the dimensionality of the data space can be significantly
reduced, and the WF/CR algorithm can be easily formulated in the reduced data space.   A
shortcoming of the present work is the neglect of redshift distortions. However, these can
be handled by using correlation matrices of Eq. 2, properly expressed in terms of
redshift space dynamical variables. The general formalism of evaluating the  redshift
space variables, within the linear theory, was outlined by Zaroubi and Hoffman (1996).

At the time this paper had been originally submitted Kolatt \etal\ (1996)
reported on  a similar project of NLCR of the \iras\ 1.2Jy catalog. Their procedure differs
from the present one mainly in not distinguishing between the low resolution (data) and high
resolution (realizations). The input data is smoothed on the $500 \, km/s$ scale and is
heavily dominated by the noise, which is `removed' by a power preserving filter (PPF)
which is a  modified WF. The PPF is designed to preserve the power, regardless
of the noise level. The resulting estimator is therefore more dominated by the noise and
less by the \prior\ model compared to our method.  Next, the resulting  $500 \, km/s$ PPFed
field is   taken from the quasi- to the linear regime by the Nusser and Dekel (1992) `time
machine', where a further Gaussianization is applied to the linear field. Small scale
structure is then added to the linearized and Gaussianized $500 \, km/s$-smoothed field by
constrained realizations. In comparison to the Kolatt \etal\   the present algorithm is more
rigorous and consistent with the theoretical framework. Our procedure involves only one
{\it ad hoc} step, namely the linearization procedure, where Kolatt \etal\ used the PPF,
linearization and Gaussianization to obtain the $500 \, km/s$-smoothed field. Yet, Kolatt
\etal\ used a better and more consistent linearization procedure, which can replace the one
used here. In spite of the very different approaches it seems that
both methods
yield similar results and are equally efficient. Detailed comparisons against N-body
simulations and mock catalogs are needed to judge the merits of each method.

\vjec

\bigskip
\cntl{\bf   Acknowledgments}
\bigskip

The members of the \iras\ collaboration are gratefully acknowledged for their help with the
\iras\ data base. We have benefited from many stimulating discussions with L. da Costa, A.
Dekel, O. Lahav and S. Zaroubi. This work is  supported in part by the
US-Israel Binational Science Foundation grant   94-00185  and by the
Israel Science Foundation grant  590/94.

\vjec

\bigskip
\cntl{\bf References}
\bigskip

\prref Bunn, E., Fisher, K.B., Hoffman, Y., Lahav, O., Silk, J., \&
Zaroubi, S. 1994, \ApJLet , {\bf 432}, L75.

\prref Bunn, E., Hoffman, Y.,  \& Silk, J.,  1996, \apj , {\bf 464}, 1.

\prref Dekel, A., 1994, {\it Ann. Rev. Astron. Astrophys.}, {\bf 32}, 371.

\prref Fisher, K.B., 1992, {\it private communication}.

\prref Fisher, K.B., Davis, M., Strauss, M.A., Yahil, A., Huchra, J.P., 1993,
\apj, {\bf 402}, 42.

\prref Fisher, K.B., Davis, M., Strauss, M.A., Yahil, A., Huchra, J.P., 1994,
\mnras, {\bf 267}, 927.

\prref Fisher, K.B., Lahav, O., Hoffman, Y., Lynden-Bell, D. \& Zaroubi,
S., 1995a,  \MNRAS, {\bf 272}, 885.

\prref Fisher, K.B., Huchra, J.P., Davis, M., Strauss, M.A., Yahil, A., \&
Schelegel, D., 1995b, \ApJS, {\bf 100}, 69.

\prref Ganon, G., \& Hoffman, Y., 1993, \ApJLet, {\bf 415}, L5.

\prref Hoffman, Y. 1993, Proc. of the
${\rm 9^{\rm th}}$ IAP Conference on {\it
Cosmic Velocity Fields}, eds. F. Bouchet and M. Lachi\'eze-Rey,
(Gif-sur-Yvette Cedex: Editions Fronti\'eres), p. 357.

\prref Hoffman, Y. 1994, in `{\it Unveiling
Large Scale Structures Behind the Milky-Way}', eds. C. Balkowski and R.C.
Kraan-Korteweg, PASP conference series.

\prref Hoffman, Y. \& Ribak, E. 1991, \ApJLet, {\bf 380}, L5.

\prref Kaiser, N. 1986, \mnras, {\bf 227}, 1.

\prref Kolatt, T., Dekel, A., Ganon, G., \& Willick, J.A., 1996, \apj, {\bf
458}, 419.

\prref Kraan-Korteweg, R.C., Woudt, P.A., \& Henning, P.A., 1996, (astro-ph/9611099).

\prref Lahav, O. 1993, Proc. of the ${\rm 9^{\rm th}}$ IAP Conference
on {\it Cosmic Velocity Fields}, eds. F. Bouchet and M.
Lachi\'eze-Rey,(Gif-sur-Yvette Cedex: Editions Fronti\'eres) p. 205.

\prref Lahav, O. 1994, in `{\it Unveiling
Large Scale Structures Behind the Milky-Way}', eds. C. Balkowski and R.C. Kraan-Kortew
eg, PASP conference series.

\prref Lahav, O., Fisher, K.B., Hoffman, Y., Scharf, C.A., \& Zaroubi,
S. 1994, \ApJLet, {\bf 423}, L93.

\prref Nusser, A., Dekel, A., Bertschinger, E., \& Blumenthal, G.R. 1991, \apj, {\bf 379}, 6.

\prref Nusser, A. and Dekel, A., 1992, \apj, {\bf 391}, 443.

\prref Padmanabhan, T., 1993,  Structure Formation in the Universe, (Cambridge:
Cambridge University Press).

\prref Peebles, P.J.E. 1980, The Large-Scale Structure of
the Universe, (Princeton: Princeton University Press).

\prref Press, W.H., Teukolsky, S.A., Vetterling, W.T., \& Flannery, B.P.
1992, Numerical Recipes (Second Edition) (Cambridge: Cambridge
University Press).

\prref Saunders, W., Rowan-Robinson, M., Lawrence, A., Efstathiou, G.,
Kaiser, N., Ellis, R., \& Frenk, C.S., 1990, \mnras, {\bf 242}, 318.

\prref Scherrer, R.J., \& Bertschinger, E., 1991, \apj, {\bf 381}, 349.

\prref Strauss, M.A., \& Willick, J.A., 1995, Phys. Rep., {\bf 261}, 271.

\prref Tegmark, M., Taylor, A.N., \& Heavens, A.F., 1996 (preprint, astro-ph/9603021).

\prref Vogeley, M.S. \& Szalay, A.S., 1996 (preprint, astro-ph/9601185).

\prref Webster, A.M., Lahav, O., \& Fisher, K.B., 1996 (preprint, astro-ph/9608021).

\prref  Wiener, N. 1949, in {\it Extrapolation and Smoothing of Stationary
Time Series}, (New York: Wiley).

\prref  Willick, J.A., Courteau, S., Faber, S.M., Burstein, D., \& Dekel. A., 1995,
\apj, {\bf 446}, 12.

\prref Yahil, A., Strauss, M.A., Davis, M., Huchra, J.P., 1991, \apj, {\bf
372}, 380.

\prref Zaroubi, S., 1995, Proceedings of the XXXth Moriond Meeting ``Clustering in the
Universe'' (in press).

\prref Zaroubi, S. and Hoffman, Y, 1996, {\it Ap. J} {\bf 462}, 25.

\prref Zaroubi, S., Hoffman, Y.,  Fisher, K.B., \& S. Lahav, O., 1995, \apj,
{\bf 449}, 446 (ZHFL).

\prref Zaroubi, S., Hoffman, Y., \& Dekel, A., 1996, preprint.

\vjec

\bye

\end